\documentclass[journal]{IEEEtran}
\ifCLASSINFOpdf
\else
\fi
\usepackage{amsmath,amssymb,amsfonts}
\usepackage{graphicx}
\usepackage{xcolor}
\usepackage[hyphens]{url}
\usepackage{color, soul}
\usepackage{float}

\usepackage{caption}
\begin{document}
%
\title{A Lane Change Assistance System\\Based on Prediction of Driver Intention}
%
%
%
\author{Foghor Tanshi, 
Dirk Söffker,\IEEEmembership{Member, IEEE}
\thanks{M. Shell was with the Department
of Electrical and Computer Engineering, Georgia Institute of Technology, Atlanta,
GA, 30332 USA e-mail: (see http://www.michaelshell.org/contact.html).}
\thanks{J. Doe and J. Doe are with Anonymous University.}
\thanks{Manuscript received April 19, 2005; revised August 26, 2015.}}

%
%

\markboth{Journal of \LaTeX\ Class Files,~Vol.~14, No.~8, August~2015}%
{Shell \MakeLowercase{\textit{et al.}}: Bare Demo of IEEEtran.cls for IEEE Journals}
%



\maketitle

\begin{abstract}
Lane change assistance system increase safety by providing warnings and other stability assistance to drivers to avert traffic dangers. In this contribution, lane change intention recognition was performed and applied to generate warnings for drivers to avoid eminent collision. Previous studies have not yet integrated driver's intended lane change actions as an input for determining when to warn drivers about eminent traffic dangers. Thus, if a driver's intended action may result in a collision, the driver should be warned in advance. In this contribution, lane change to left and right and lane keeping intentions were utilized to warn drivers of potential collision using an audio visual interface. The results indicate reduced risk of collision during lane change to left and right except lane keeping maneuvers. Moreover several participant feedback indicate an increased need for improved warnings by additional situational analysis that anticipate other vehicle behaviors such as intended lane changes. 
\end{abstract}


%
\IEEEpeerreviewmaketitle

\section{Introduction}
%
%
%
%


\IEEEPARstart{S}{tatistics} of traffic accidents show that most accidents are caused by driver errors, distraction and fatigue~\cite{Ryder:2017}\cite{NSC:2018,NHTSA:2019}. Drivers account for 94 \% of vehicle crashes and lane change as well as merging accounts for approximately 9 \% of fatal accidents~\cite{NHTSA:2015}. Accordingly, lane change assistance systems are integrated into vehicles to increase safety by providing warnings, information, avoiding lane departures and stabilizing lane changes. 

Lane change assistance (LCA) systems enable drivers to safely change lanes without colliding with vehicles on the adjacent lane. They integrate various functionalities such as blind spot, frontal and rear collision warnings that are displayed on the dashboard. To improve the utility of LCA, recent studies have focused on recognising driver intention, improving interface design, and improving driver trust as further discussed below.

\begin{figure}[h!]
	\centering
	\includegraphics[width=0.96\columnwidth]{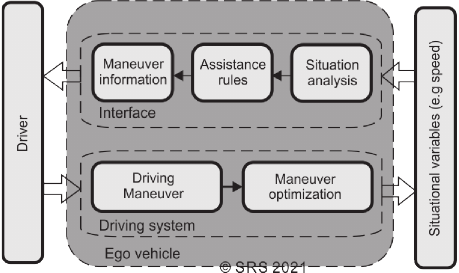}
	\caption{Online driver assistance system}~\label{fig:DVI_loop}
\end{figure}

Recognized driver intention can be used as input to evaluate whether driver's intended action is safe in each context. If situations may become unsafe a corresponding maneuver suggestion or a warning can be given to enable safe actions as illustrated in Fig. \ref{fig:DVI_loop}.  Variations in the suggestions could be adapted to the driver's situational needs. This can also be used to determine when to provide maneuver assistance that optimizes driving inputs such as steering and braking assistance \cite{Khan:2019}.

In~\cite{Yan:2017,Fu:2020}, different speed thresholds in which drivers assume lane change warnings are necessary based on individual perception were identified. The drivers indicated increased trust for the system because the warnings were adapted to their individual's thresholds. However, the studied scenarios only integrated front right and left rear approaching vehicles. In addition, intention recognition and directional information as to why the lane change was unsafe were not integrated.


In~\cite{Yan_yang:2020,GUO2019326}, lane change intention recognition was applied to shared steering control and haptic guidance respectively. The steering automatically optimizes the maneuvers of the driver based on the predicted intention. The authors concluded that the systems support stable lane changes and decreases lane departure risk. However, the applied intention recognition approaches were not used to warn drivers of potential collisions. Moreover, a method for anticipating the lane change intention of surrounding vehicles resulted in decreased near collisions by approximately 30 \% using a simulation~\cite{Williams:2018}. 

Moreover, some interfaces that have been utilized for lane change warnings include audiovisual cues, augmented reality (AR) head up display (HUD), instrument cluster display, adaptive light intensity, and constant light intensity~\cite{locken:2019,KRAFT202081}. The aforementioned interfaces when utilized to indicate the direction of an eminent collision, increase driver SA but have not been combined with intention recognition. 

Driver intention has been integrated in providing haptic guidance to enable the driver complete an intended manoeuvre including lane change \cite{WangLWWJ2023}. While another study utilised gaze data-based intention recognition in a lane keeping assistant for correction of unintentional lane departure \cite{DahldCGF2023}.

Furthermore, in~\cite{Zhu:2018}, aggressive, normal and cautious lane change behaviors were classified and used to generate warnings for three participants in a driving simulator. The results indicate that warning drivers based on individual driving styles leads to reduced risk of lane change collisions. However, this included only three participants which is insufficient to conclude if the prediction of driving behavior reduces lane change accidents on a larger scale. Given the abundance of methods for prediction of driver lane change intention as reviewed in~\cite{DIAZALVAREZ2018,Xing:2019}, very few studies have integrated it and are mostly related to stability. In addition, utilizing driver intention prediction to change lanes have not been applied to warnings using an adequate number of participants.


\subsection{Research statement}
The reviewed literature reveal that there are few studies in which driver intended actions are considered in the assistance provided. It is necessary to anticipate future driver actions and warn ahead of possible danger in addition to providing assistance in the present context.  Existing systems warn drivers about the presence of surrounding traffic without also considering the intended maneuver of the driver~\cite{locken:2019,KRAFT202081,Yan:2017,Fu:2020,Yan_yang:2020,GUO2019326}. If a driver's current actions indicate an intention to change lanes when a collision is eminent, the driver should also be warned.

Furthermore, warning drivers about the presence of surrounding vehicles without consideration of their intended actions may result in drivers habitually ignoring warnings. In this contribution, it is assumed that providing warnings based on driver's intended lane change actions will increase compliance and reduce related crashes. 





 

\subsection{Outline of contribution} 
This contribution focuses on the study of lane change assistance warnings given driver predicted intention as input. The outline of this contribution includes an introduction and review of studies related to lane change assistance systems. Next, the algorithms and variables included in the study are discussed. These include prediction of driver lane change intention developed in a previous contribution~\cite{DengSTS:2020}, the driving scenario, and warning timing thresholds. Finally, the discussion of results, summary, limitation, and outlook are provided.

\section{Methods}
Although driver intention is an internal state of mind that cannot be read, it is possible to utilize ego vehicle trajectory measurements as inputs to machine learning algorithms to predict lane changes. These algorithms include Artificial Neural Networks (ANN) \cite{c1}, Dynamic Bayesian Networks (DBN) \cite{c2}, Support Vector Machines (SVM) \cite{YDou16},Hidden Markov Models (HMM) \cite{DengWangSoe18}, and Random Forest (RF) \cite{WCao17} which may also be combined with Fuzzy Logic (FL) \cite{c4}. 

\begin{figure}
\centering
\includegraphics[width=0.9\columnwidth]{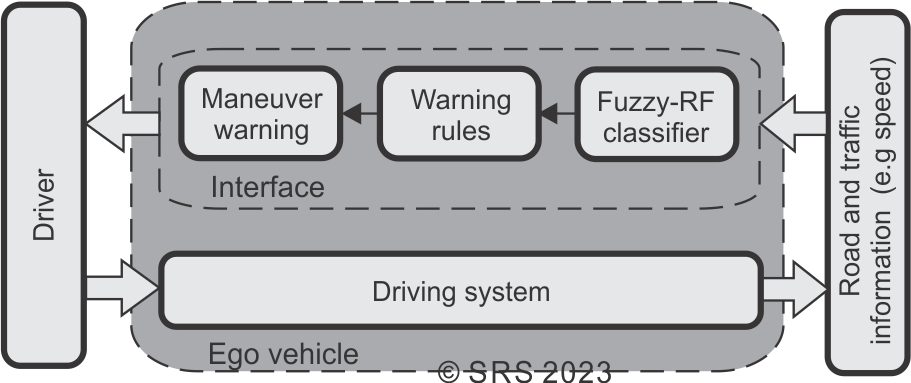}
\caption{Intention-based lane change assistance (cf.~\cite{DengSTS:2020})}~\label{fig:RF_IR_assistance_loop} 
\end{figure}

A comparison of HMM, SVM, Convolutional Neural Networks (CNN), and RF in previous contributions \cite{DengWangSoeJ1819,DengSTS:2020} indicates that the performance of RF algorithm especially in combination with fuzzy logic is the best with an accuracy of over 99 \%. To integrate driver intention to change lanes in a lane change assistance system, a fuzzy-random forest classifier from a previous contribution was utilized~\cite{DengSTS:2020}. The ADS integrating driver lane change intention is illustrated in Figure~\ref{fig:RF_IR_assistance_loop}. It integrates an individualized lane change intention recognition approach with which collision warnings were generated as further explained below. 



\subsection{Individualized lane change intention recognition}\label{IRS}

\subsubsection{Fuzzy Logic}
Fuzzy Logic (FL) is a popular approach used for modeling vagueness introducing many-valued logic.
The structure of FL is easy to interpret by using IF-THEN rules. 
The FL approach is considered an extension of Boolean logic,  it is based on fuzzy sets and allows modelling of the truth of statements continuously between true (one) and false (zero) using membership functions \cite{c4}. 
Common fuzzy sets are based on triangular, trapezoidal, or Gaussian membership functions \cite{JZhao02}. In this contribution, trapezoidal membership function is used to convert the signal data to membership degrees, the output of the FL is a vector which contain all membership degrees of signals.

\begin{figure}[h]
	\centering
	\includegraphics[width=0.7\linewidth]{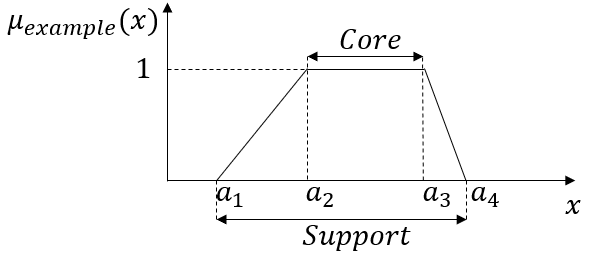}
	\caption{Trapezoidal membership function defined by core ($a_2$, $a_3$) and support ($a_1$, $a_4$) parameters}
	\label{FL_example}
\end{figure}

As shown in Fig.~\ref{FL_example} core and support parameters of each membership function (MF) are unknown and defined as design parameters, which can be determined through a fuzzy density clustering method \cite{Derbel08} \cite{Ulutagay08} called ``Fuzzy Neighborhood density-based spatial clustering of applications with noise (FN-DBSCAN)''. Therefore, the membership functions for each variable of data set can be generated automatically.

\subsubsection{Random Forest}
Random Forest (RF) is an extension of decision tree method and is used to solve classification or regression problems\cite{Bre01}. A decision tree poses a series of selection problems, and each final answer to these questions is represented by leaves. The structure of a decision tree is divided into several stages. Each non-leaf node represents a question that needs to be answered by making a decision between two or more selections. After each selection, the question of the next node becomes more specific. This process is considered as feature extraction, which are evaluated by each node and passed to one branch until finally the level is reached and thus a classification is determined. In this contribution, the driving intentions namely lane change to left (LCL), lane change to right (LCR) and lane keeping (LK) make up the three types of leaves of the decision trees.

The RF algorithm  contains a set of randomized decision trees, all the decision trees are independent from each other. Each decision tree is trained by a randomly selected Bootstrap sample \cite{BL96}, which is generated from the training data set with replacement. After these decision trees are generated, the output result of the RF is obtained through the voting results of all relevant decision trees.

\subsubsection{Labelling and training of Fuzzy-random forest (RF) classifier} 
\begin{figure}
\centering
\includegraphics[width=1.0\linewidth]{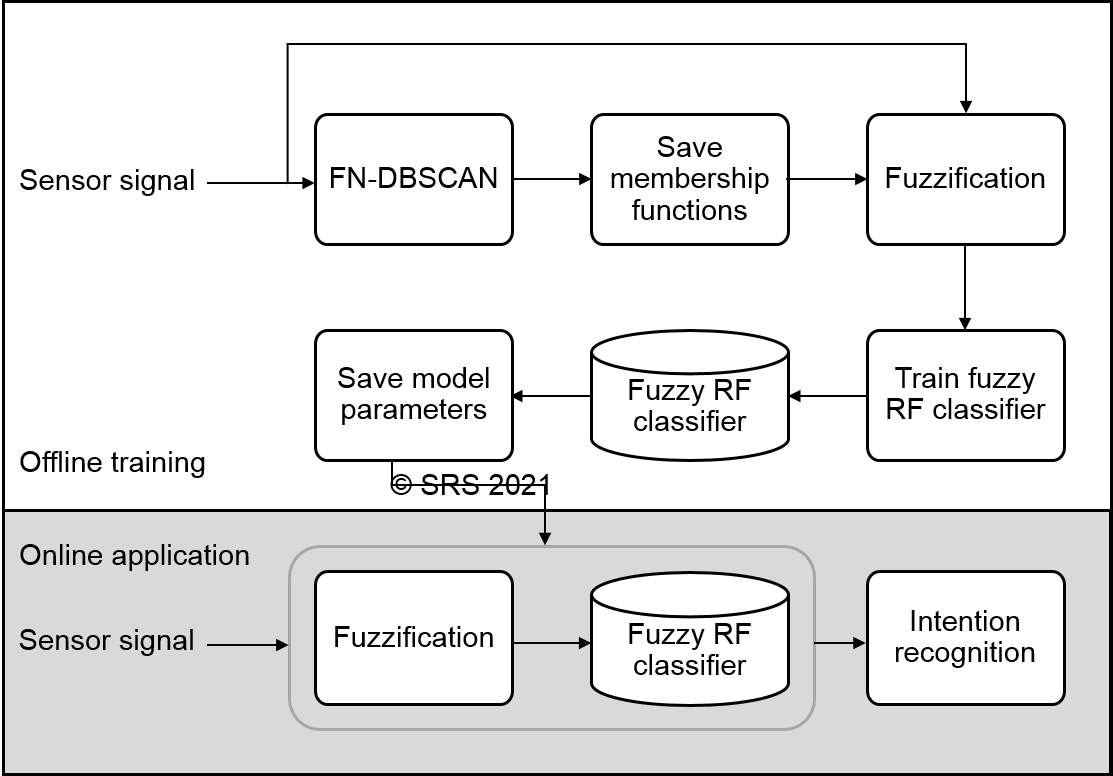}
\caption{Offline training and on-line fuzzy-RF
model adapted from~\cite{DengSTS:2020}}
\label{FL_RF}
\end{figure}

The driving intention recognition model based on fuzzy-RF is shown in Fig.~\ref{FL_RF}. It consists of off-line training and on-line test phases, which are described below.

The lane change intention recognition applied in this contribution utilizes an online Fuzzy-Random Forest (fuzzy-RF) approach. Membership functions quantify the input signal data into fuzzy sets for RF training. Three different driving maneuvers namely lane keep (LK), lane change to Left (LCL), and lane change to right (LCR) are modeled as outputs~\cite{DengSTS:2020}. 

The fuzzy-RF classifier utilizes 24 variables that affect drivers' decisions as input for training as detailed in~\ref{F_inpute}. These include the states of the ego vehicle (lane number, indicator, gearbox mode, speed, acceleration, steering wheel angle, and heading angle) and information about surrounding vehicles (TTC and distance to ego vehicle) in all six directions of a unidirectional three-lane highway. The surrounding vehicle directions relative to the ego vehicle include front (f), back (b), front left (fl), back left (bl), front right (fr), back right (br). 

\begin{table}[t]
	\centering
	\caption{Descriptions of selected features reused from~\cite{DengSTS:2020}}\label{F_inpute}
	\renewcommand \arraystretch{1.0}
	\begin{tabular}{|p{1.1cm}|p{3cm}| p{1.4cm}| p{0.5cm}|p{0.7cm}|} \hline	
		\  Input  &  Definition  &  Range & Unit & Data type \\	
		\hline
		\multicolumn{5}{|c|}{Category \#1}   \\	\hline
		\  ${v}_{ego}$  & Velocity of ego vehicle & [0 220] & km/h & Real \\ \hline
		
		\multicolumn{5}{|c|}{Category \#2}   \\	\hline
		\  ${v}_{f}$  & Velocity of vehicle in front & [0 220] & km/h & Real \\
		\ ${v}_{fl}$  & Velocity of vehicle in left-front & [0 220] & km/h & Real  \\
		\ ${v}_{fr}$  & Velocity of vehicle in right-front & [0 220] &  km/h & Real  \\
		\ ${v}_{bl}$  & Velocity of vehicle left-behind & [0 220] & km/h & Real \\
		\ ${v}_{br}$  & Velocity of vehicle right-behind & [0 220] & km/h & Real  \\
		\ ${v}_{b}$  & Velocity of vehicle behind & [0 220] & km/h & Real  \\
		\hline

		\multicolumn{5}{|c|}{Category \#3}   \\	\hline
		\ ${d}_{f}$  & Distance to vehicle in front & [0 250] & m & Real  \\
		\ ${d}_{fl}$  & Distance to vehicle in left-front & [0 250] & m & Real  \\
		\ ${d}_{fr}$  & Distance to vehicle in right-front & [0 250] & m & Real  \\
		\ ${d}_{bl}$  & Distance to vehicle left-behind & [0 250] & m & Real  \\
		\ ${d}_{br}$  & Distance to vehicle right-behind & [0 250] & m & Real  \\
		\ ${d}_{b}$  & Distance to vehicle behind & [0 250] &  m & Real  \\
		\hline
		
		\multicolumn{5}{|c|}{Category \#4}   \\	\hline
		\ ${TTC}_{f}$ & TTC to vehicle in front  & [0 12] & s & Real \\
		\ ${TTC}_{fl}$ & TTC to vehicle in left-front  & [0 12] & s & Real \\
		\ ${TTC}_{fr}$ & TTC to vehicle in right-front  & [0 12] & s & Real \\
		\  ${TTC}_{bl}$ & TTC to vehicle left-behind  & [0 12] & s & Real \\
		\ ${TTC}_{br}$ & TTC to vehicle right-behind  & [0 12] & s & Real \\
		\ ${TTC}_{b}$ & TTC to vehicle behind  & [0 12] &  s & Real \\	
		\hline
		
		\multicolumn{5}{|c|}{Category \#5}   \\	\hline
		\  $ \alpha $ & Heading angle of ego-vehicle  &  [-3.14 3.14] & rad & Real\\
		\hline
		
		\multicolumn{5}{|c|}{Category \#6}   \\	\hline
		\  ${S}$ & Steering wheel angle  & [-3.14 3.14] & rad & Real\\
		\hline

		\multicolumn{5}{|c|}{Category \#7: Fuzzification is not necessary}   \\	\hline
		\  $ Ln $ & Current lane number & [$1,\ 2$] & - & Integer\\	
		\ ${I}$ & Indicator  & [$0,\ 1,\ 2$] & - & Integer\\
		\ ${G}$ & Gearbox  & [$1,... 5$] & - & Integer\\

		\hline			
	\end{tabular}
\end{table}

\begin{figure}
	\centering
	\includegraphics[width=1.0\linewidth]{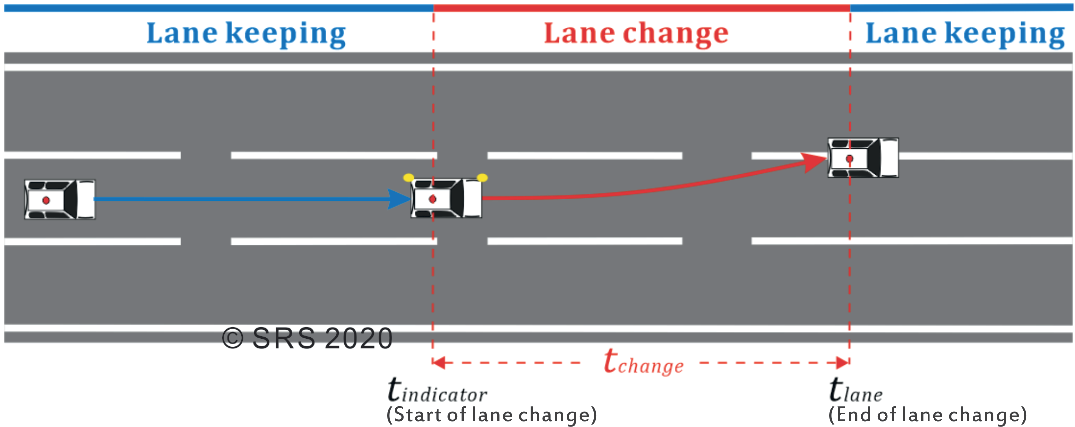}
	\caption{Intention to eventual lane change~\cite{DengSTS:2020,DengWangSoeJ1819}}
	\label{Lanechange}
\end{figure}

To automatically label the data as intended driving maneuvers (LK, LCL, and LCR), the current lane $i$ is first determined by the position of the center point of the ego vehicle. The beginning of an intended lane change maneuver is the moment the driver turns on the indicator and it is denoted as ${t}_{indicator}$~\cite{DengWangSoeJ1819}. The lane change at time $t_{lane}$ is recognized when the value of lane $i$ changes e.g from lane 1 to lane 2. The variable $t_{lane}$ denotes the end time of the lane change as illustrated in Figure~\ref{Lanechange}.  The interval between the beginning of the lane change (${t}_{indicator}$) and the end of the lane change (${t}_{lane}$) is the total time (lane change time) denoting the lane change which is defined as ${t}_{change}$. When the drivers make lanes changes without turning on the indicator, the average value of 2.5 s is taken as ${t}_{change}$ which is the lane change time.

Furthermore, the decision to use ${t}_{change}$ as 2.5 s was made based on comparison from previous contributions~\cite{DengWangSoeJ1819,DengSTS:2020}. In~\cite{DengSTS:2020} the relative angles of 0.5 $^{\circ}$, 0.2 $^{\circ}$, and 0.5 $^{\circ}$ between the road and the ego-vehicle was used to determine the start time of the lane change intention whenever a driver does not turn on the indicator. The previous results which compared ${t}_{change}$ of 3 s, 2.5 s, and 2 s as well as relative angles of 0.05 $^{\circ}$, 0.2 $^{\circ}$, and 0.5 $^{\circ}$ indicate that ${t}_{change}$ of 2.5 s results in an accuracy of more than 99.6 \%~\cite{DengWangSoeJ1819,DengSTS:2020}. 


The intention recognition process consists of an individualized offline training and an online application of the output model for assistance as illustrated in Figure~\ref{FL_RF}. The participants were invited to an initial appointment to provide data by driving manually for about 25 to 35 minutes which was used to generate membership functions and for training an individualized fuzzy-RF model. Finally, the models for each driver were saved separately for online lane change intention recognition during the experiment. Thus, the offline training was performed individually for each driver.

\subsection{Scenario:} 
\begin{figure}
	\centering
	\includegraphics[width=0.95\linewidth]{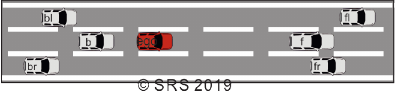}
	\caption{Interacting vehicles with Ego vehicle on the highway}
	\label{fig:int_vec}
\end{figure}

The two scenarios (S1nad S2) utilized in the initial training data collection and online lane change intention assistance system phases in this study are two-way highways, each consisting of three lanes and automatically generated interacting vehicles as illustrated in \ref{fig:int_vec}. In the training data collection phase, S1 a bright daytime weather condition (without rain, snow, storm etc.) was implemented in the scenario. The interacting vehicles had different driving behaviors namely cautious, normal, and aggressive. Aggressive vehicles performed short overtaking maneuvers using shorter TTC of 2.2 s to 3 s to front vehicles. Normal and cautious vehicles performed longer overtaking maneuvers using longer TTC to front vehicles. The percentage distribution of cautious, normal, and aggressive vehicles were 20 \%  60 \% and 20 \% respectively.

In the online lane change intention assistance system phase, the scenario complexity was increased in S2 by integration of a Light foggy daytime weather condition (without rain, snow, storm etc.). In addition, the percentage distribution of cautious, normal, and aggressive vehicles were 20 \%  30 \% and 50 \% respectively. In this contribution, it is assumed that inserting more aggressive driving behaviors are necessary to subject the participants to conditions that might require warnings. Having significant warnings is necessary for the evaluation of the drivers' compliance to the warnings.

\subsection{Warning timing thresholds and related imagery} Time-to-collision (TTC) was utilized to determine the warning thresholds together with intention recognition. This is the time required by two objects to collide if the current longitudinal speed is maintained. The TTC should be at least 3 s between lead and following vehicles for safe lane changes~\cite{WANG2019127}. Accordingly, 4 s and 3 s TTC was selected as the warning threshold for front and back vehicles whenever a lane keep intention was recognized. The additional 1 s for the front vehicle was to ensure that the drivers had sufficient time to respond after receiving warnings. To determine the warning thresholds for lane changes, the average lane change duration of 2.5 s on highways was added to the aforementioned 3 s required between front and back vehicles~\cite{WANG2019127}. This average lane change duration was added to the TTC of adjacent lead and following vehicles to obtain a 5 s warning threshold for lane changes.

\begin{figure}
	\centering
	\includegraphics[width=0.55\linewidth]{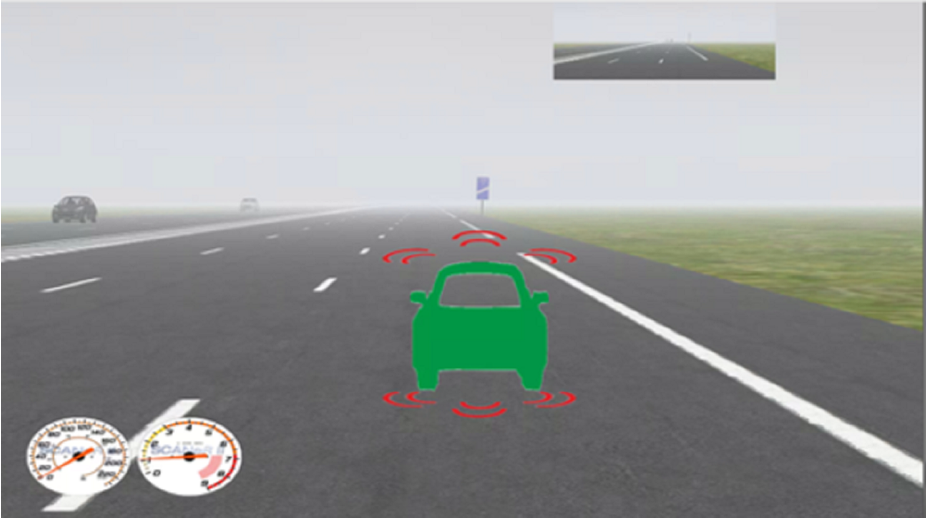}
	\caption{Intention collision warning imagery on the HUD}
	\label{fig:warning_imagery}
\end{figure}

\begin{figure}
	\centering
	\includegraphics[width=1.0\linewidth]{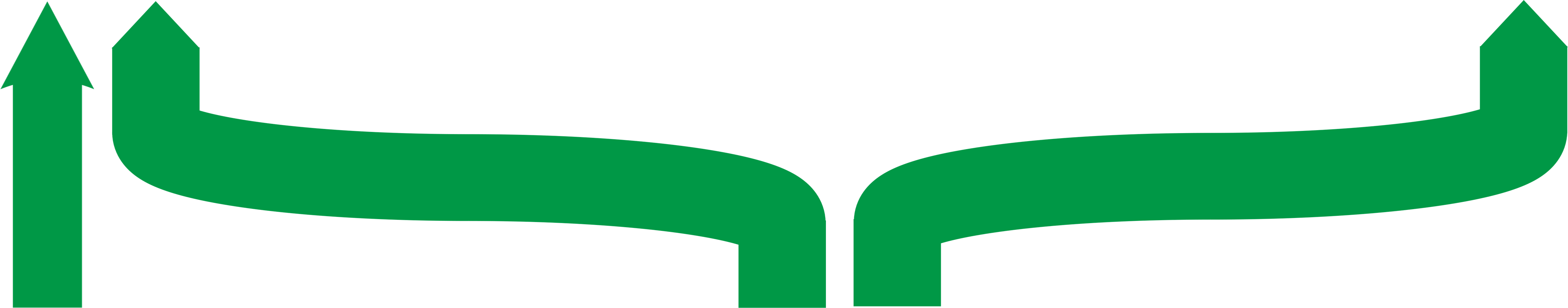}
	\caption{Intention approval imagery for LK, LCL, and LCR respectively on the HUD}
	\label{fig:approval_imagery}
\end{figure}

As recommended in~\cite{SAE:2002}, the opinion of experienced drivers were solicited to determine the applicable surrounding vehicles to warn the drivers about. Four experienced drivers with 10 years to 50 years of driving experience were polled. Recommendations from the experienced drivers included warning drivers about vehicles on the intended lane and then anticipatory warning about vehicles on the same lane. Altogether, six warnings with different TTC thresholds were included as illustrated in Figure~\ref{fig:warning_imagery}. These include for LK (f = 4 s, b = 3 s, fl = 3.5 s, bl = 3.5 s), LCL (f = 4.5 s, b = 4 s, fl = 5.5 s, bl = 5.5 s), and LCR (f = 4.5 s, b = 4 s, fr = 5.5 s, br = 5.5 s). In other words, if the TTC values were less than the aforementioned ones at the same time that the predicted intention was either LK, LCL, or LCR respectively, the drivers were warned. Generally, these thresholds were selected to be long enough for the drivers to respond after the warnings without being too early or too late. 

Furthermore, three images that indicate the ADS approves of a driver's intended maneuver as safe where included to encourage cautious lane changes as illustrated in Figure~\ref{fig:approval_imagery}. The timing threshold for displaying each approval imagery whenever the predicted intention occurred was obtained by adding 2 s to the corresponding warning thresholds. Thus these include for LK (f = 6 s, b = 5 s, fl = 5.5 s, bl = 5.5 s), LCL (f = 6.5 s, b = 6 s, fl = 7.5 s, bl = 7.5 s), and LCR (f = 6.5 s, b = 6 s, fl = 7.5 s, bl = 7.5 s) respectively.  

In addition, by using the recommendation of the polled experienced drivers, the warnings and approvals were displayed for 2 s. The 2 s was accepted as sufficient time for the warning display because in manual driving at high speeds, situations are highly dynamic. Otherwise, the displayed information would become obsolete in the event of a longer display time. Several applicable warnings based on the defined thresholds were displayed at the same time and accompanied by the audio expression ``collision risk". Specifying the direction of the risk enables warnings to be comprehensible and avoids increased workload in reconciling view reference frame~\cite{Naujoks:2016}.


\subsection{Experimental participants and design}\label{participants}
 Drivers having different levels of experience were invited to participate in the study. A total of 44 drivers (39 males and 5 females) among whom 22 experienced the lane change assistance system with intention recognition and corresponding warnings in scenario S2. Each of the 22 drivers constituted an experimental group who had attended an initial appointment to drive in the previously described scenario S1 to provide data from which intention recognition models were trained and utilized in the assistance system to generate directional warnings as previously illustrated in Figure~\ref{fig:warning_imagery}. Directional warnings are effective in different scenarios depending on driving experience and gender~\cite{YUE2021336}.
 
 The remaining 22 drivers experienced scenario S2 without lane change assistance warnings and did not attend any previous appointment for generation of training data. In other words, the second set of 22 drivers constituted a control group that experienced scenario S2 without directional warnings.
 
 All the drivers held valid driver's licenses, 10 of them have previously used vehicles equipped with one or more ADS features, 11 of them had previously experienced driving simulators and additional descriptive statistics obtained from the pre-questionnaire include age [Yrs] (mean = 27.2, STD = 3.5, min = 20, max = 35.1), driving experience [Yrs]  (mean = 7.5, STD = 4.2, min = 0.4, max = 17.8), and driving experience [km/wk]  (mean = 256.7, STD = 1035.3, min = 5.0, max = 7000). 

As their reward, the participants either received 15 EUR or attended a three-hour time management seminar. In accordance with the relevant ethics rules, participants signed a participation consent declaration. The participants were also informed that their participation was voluntary and that they were free to discontinue the experiment if desired.

\subsection{Experimental environment}
\begin{figure}
\centering
  \includegraphics[width=1.0\columnwidth]{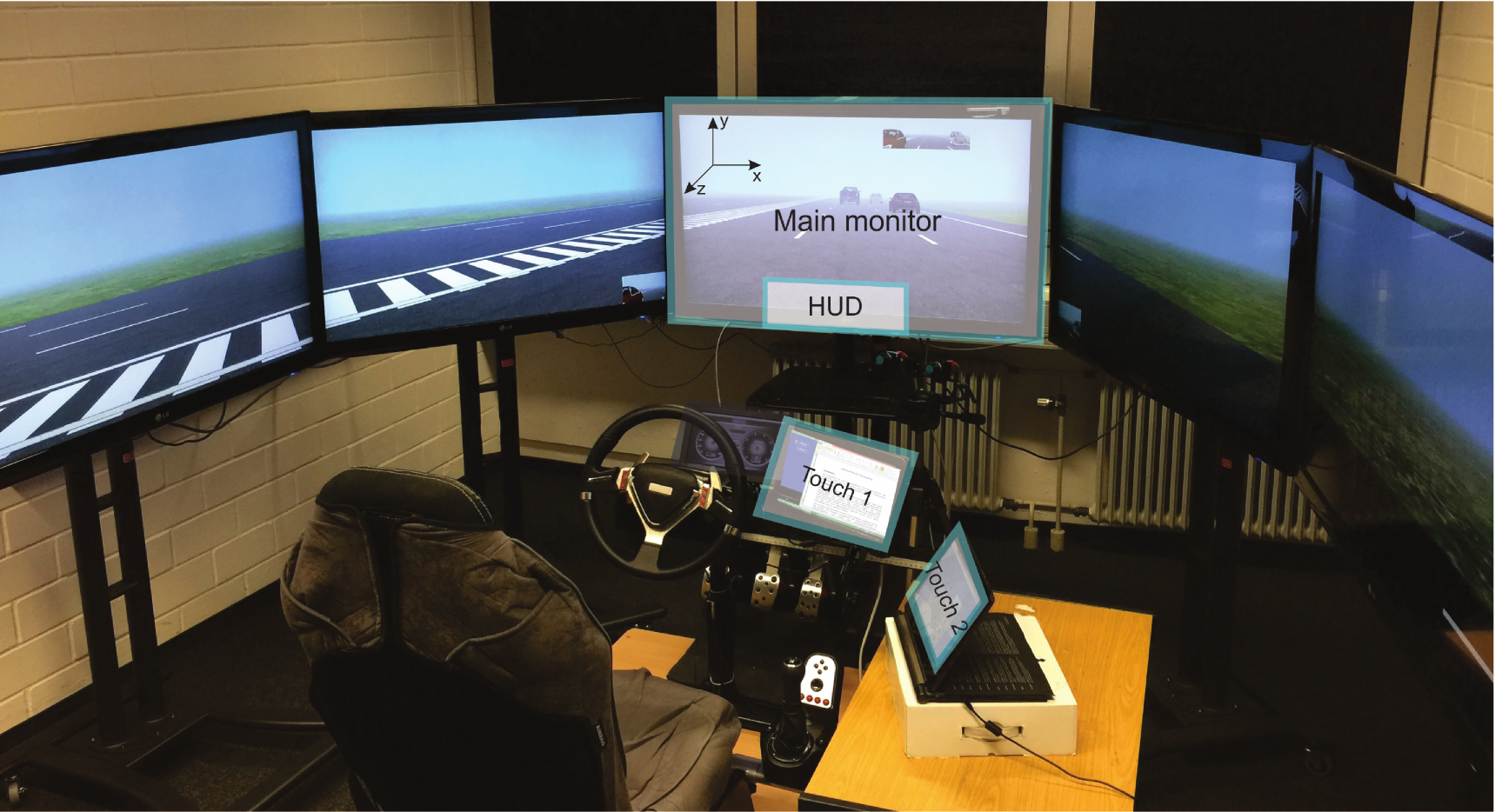}
  \caption{Driving simulator lab., Chair SRS, U DuE, Germany}~\label{fig:DS}
\end{figure}

The scenarios were implemented in SCANeR\textsuperscript{TM} studio (a professional driving simulator software by AVSimulation). The data acquiring frequency of SCANeR\textsuperscript{TM} studio is 20 Hz. The driving simulator setup includes five displays that provide 270 \textsuperscript{0} field of view, a fixed-base driver seat, steering wheel, clutch, brake, and accelerator pedals as displayed in Figure~\ref{fig:DS}. A rear view mirror and two side mirrors were displayed on the appropriate positions of the monitors. A control pad (Touch 1) displays the driving modes on the screen. In addition, another display (Touch 2) was used by all the participants to complete the questionnaires.

\subsection{Experimental procedure}
As previously stated, the study was performed in two stages. In the first stage, the experimental group were invited to an initial data collection appointment that lasted approximately one hour. The participants were given an introduction to the goal of the experiment and allowed to test drive the simulator for about 10 minutes for adaptation. Then they were allowed to drive in S1 for approximately 30 minutes from which data was obtained to train the individualized Fuzzy-RF intention recognition model.

\begin{figure}
	\centering
	\includegraphics[width=0.75\columnwidth]{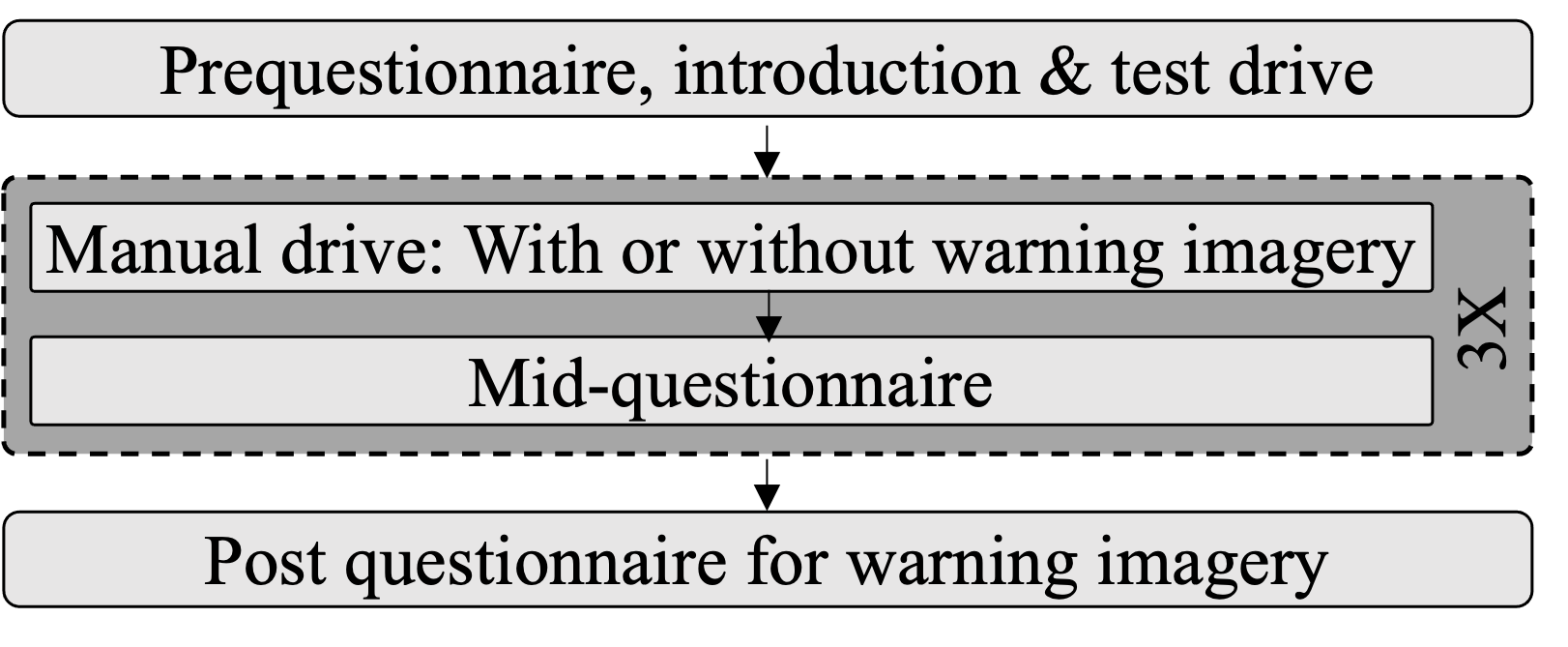}
	\caption{Experimental procedure}~\label{fig:exp_procedure}
\end{figure}

Each second stage laboratory appointment lasted approximately two hours and the procedure is summarized in Figure~\ref{fig:exp_procedure}. First, the participants filled a pre-questionnaire about their driving experience previously detailed in Section~\ref{participants}. Afterwards, the participants received an introduction to the simulator and the procedure as well as the goal of the study for approximately 10 minutes. The visual directional warning imagery as well as the audio sound ``collision risk" with which the warnings were given was also shown to the experimental group participants.

Afterwards, the participants performed a test drive for approximately 10 minutes to adapt to the driving simulator. Finally, they performed three experimental drive procedures using S2 including filling questionnaires. After each drive, the experimental group participants were asked to fill a mid-questionnaire about the assistance system and the scenario. Each drive procedure lasted approximately 25 to 30 minutes with a few minutes break in between. In addition, the experimental group also filled a post questionnaire at the end of all the three drive procedures.


\subsection{Data measures and analysis approach}\label{analysis_approach}
The measured variables include time to collision and participant questionnaire ratings.

\subsubsection{Time to collision:} The number of times a participant performed each of the three maneuvers (LK, LCL, LCR) at critical TTC thresholds based on~\cite{Naujoks:2016} was compared to the number of each maneuver performed during the drive. In other words, the ratio of violations were computed for all surrounding vehicles that posed a risk of collision. Mathematically, the ratio of TTC violations is expressed as 

\begin{equation} 
\small
TTC~[s] =
\sum \frac{Exceeded~threshold~counter}{ LK,~LCL,~or~LCR~counter}.
\end{equation}

Expressing TTC this way ensured that the behavior of the participants throughout the drive was adequately represented compared to instantaneous observations using minimum or maximum TTC.

\subsubsection{Mid questionnaire:} The ratings of the experimental group on a 5-point Likert scale about helpfulness, SA increment, comprehensibility, timeliness, and increase in performance due to the ADS was adapted from~\cite{Wang:2019,Naujoks:2016}. 

\subsubsection{Post questionnaires:} The ratings of the experimental group on a 5-point Likert scale about acceptability, ease of use, usefulness, noticing danger early, increase in safety, intent to use, and endangering others due to the ADS was adapted from~\cite{Naujoks:2016}. 

The warnings were analyzed in relation to the previously defined TTC performance variable using t-test. In other words, the results of the experimental and control groups were compared to determine if the intention-based lane change assistance warnings improves driving performance. A significance level of p = 0.05 was used for the analysis. 

\section{Evaluation of results and discussion}

\subsection{Results}
\begin{figure}
	\centering
	\includegraphics[width=0.8 \columnwidth]{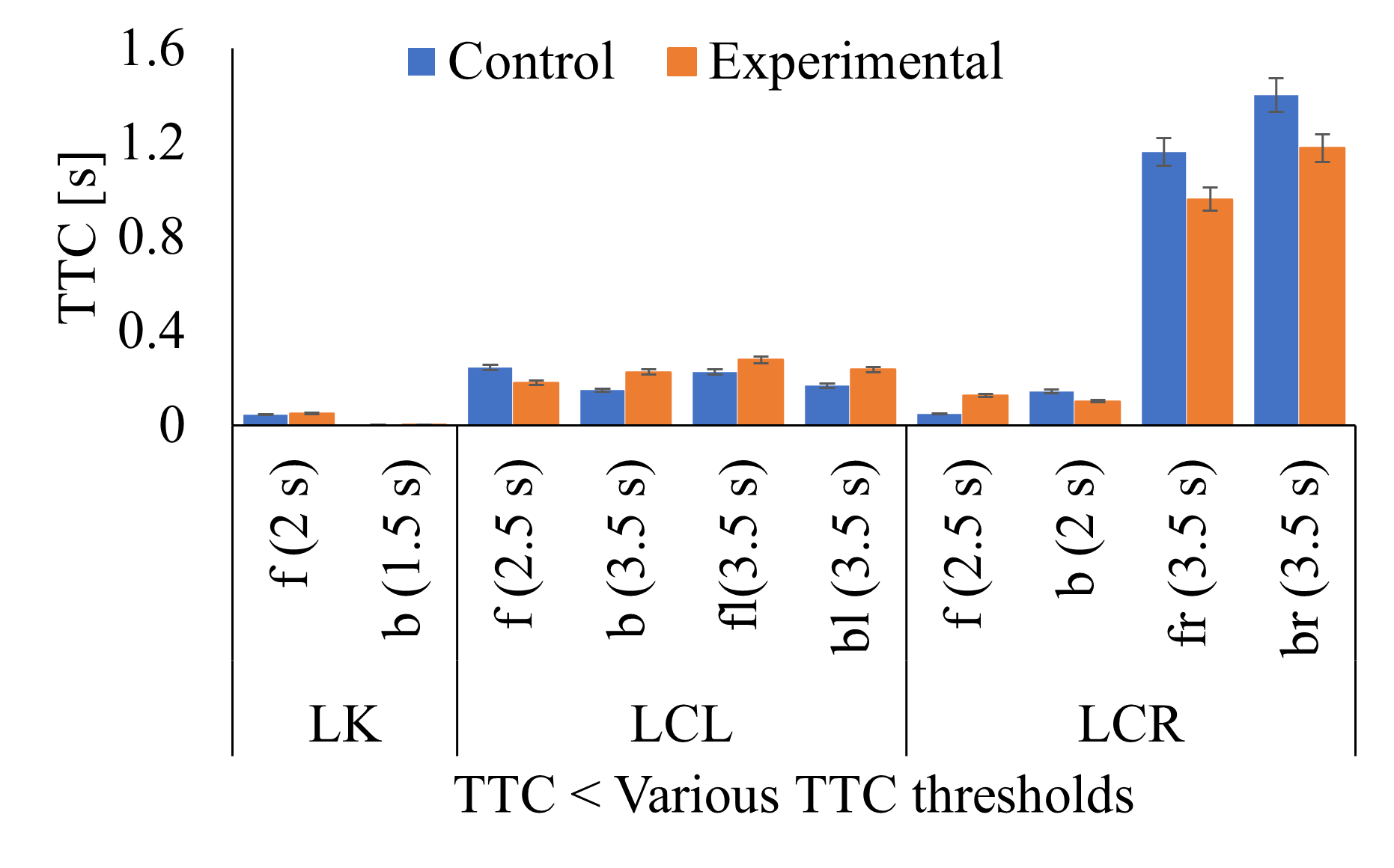}
	\caption{Performance rate at various TTC thresholds
	}~\label{fig:Intention_performance_TTC} 
\end{figure}

\begin{figure}
	\centering
	\includegraphics[width=0.8 \columnwidth]{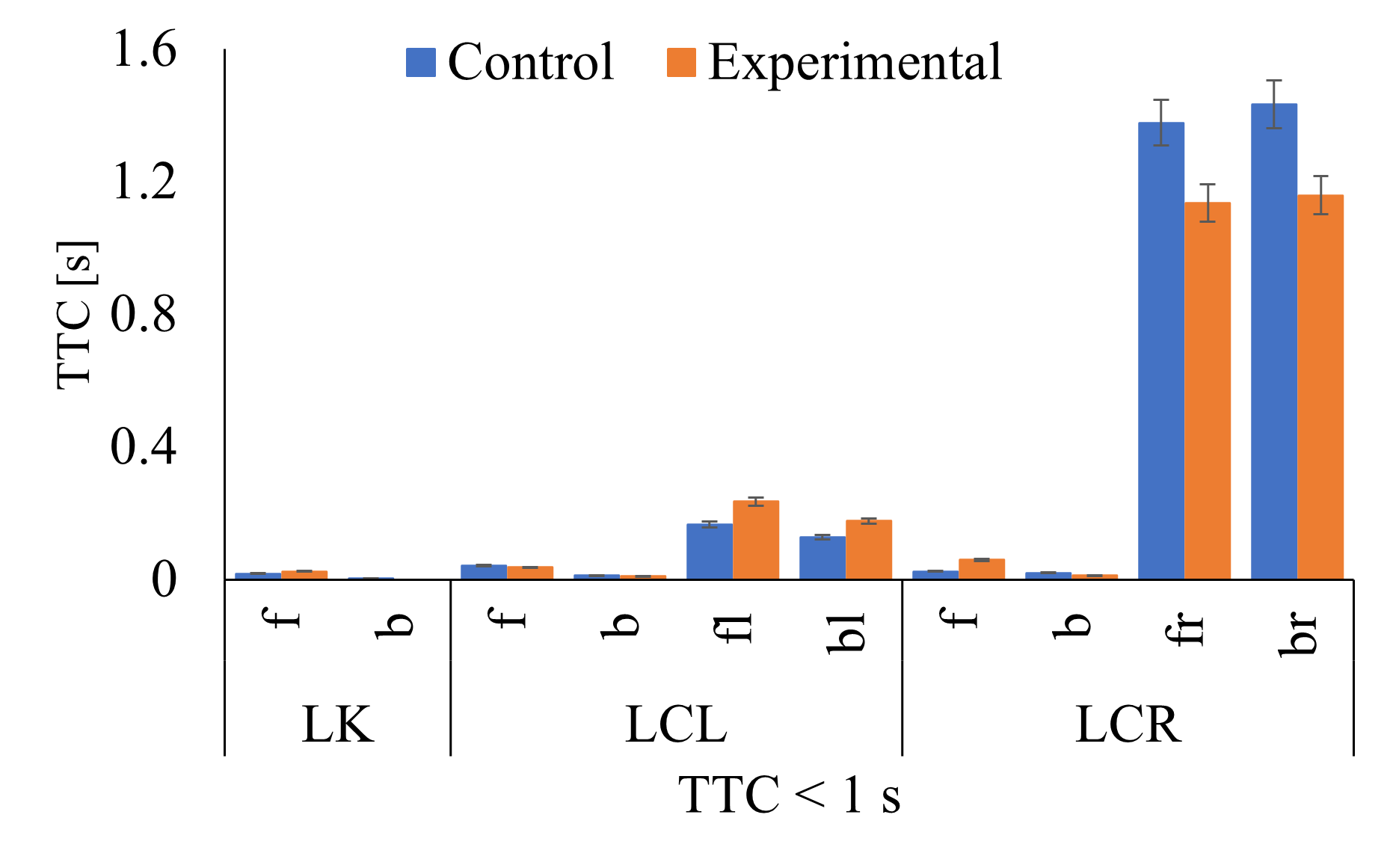}
	\caption{Near misses using a TTC $<$ 1 s
	}~\label{fig:Intention_accident_TTC} 
\end{figure}

Different TTC thresholds reduced in steps of 0.5 s from the warning thresholds were used to identify where the behavior of the experimental and control groups were significantly different in order to illustrate where drivers heeded or ignored warnings (Figure~\ref{fig:Intention_performance_TTC} and Figure~\ref{fig:Intention_accident_TTC}). In Figure~\ref{fig:Intention_performance_TTC} the threshold values evaluated are of different magnitudes. In Figure~\ref{fig:Intention_accident_TTC}, the threshold value evaluated was 1 s which indicates either accident or near misses.

 Using t-test, significantly (p $<$ 0.05) improved performance was found for f vehicles during LCL and fr and br vehicles during LCR maneuvers in the experimental group due to reduced near misses.  Most near misses (TTC $<$ 1 s) occurred during LCR maneuvers and participants are more cautious during LCL maneuvers. The performance of the experimental group participants is significantly (p $<$ 0.05) worse for b and bl vehicles during LCL and f vehicles during LCR maneuvers. This result indicates that the experimental group participants mostly ignored the warnings for b and bl vehicles during LCL and f vehicles during LCR maneuvers. The drivers may have ignored the warning due to lack of acceptance and trust in the system, personal threshold preferences or complexity of the scenario demanding more workload.


\begin{figure}
	\centering
	\includegraphics[width=0.6\columnwidth, trim={0 0 11.4cm 0}, clip]{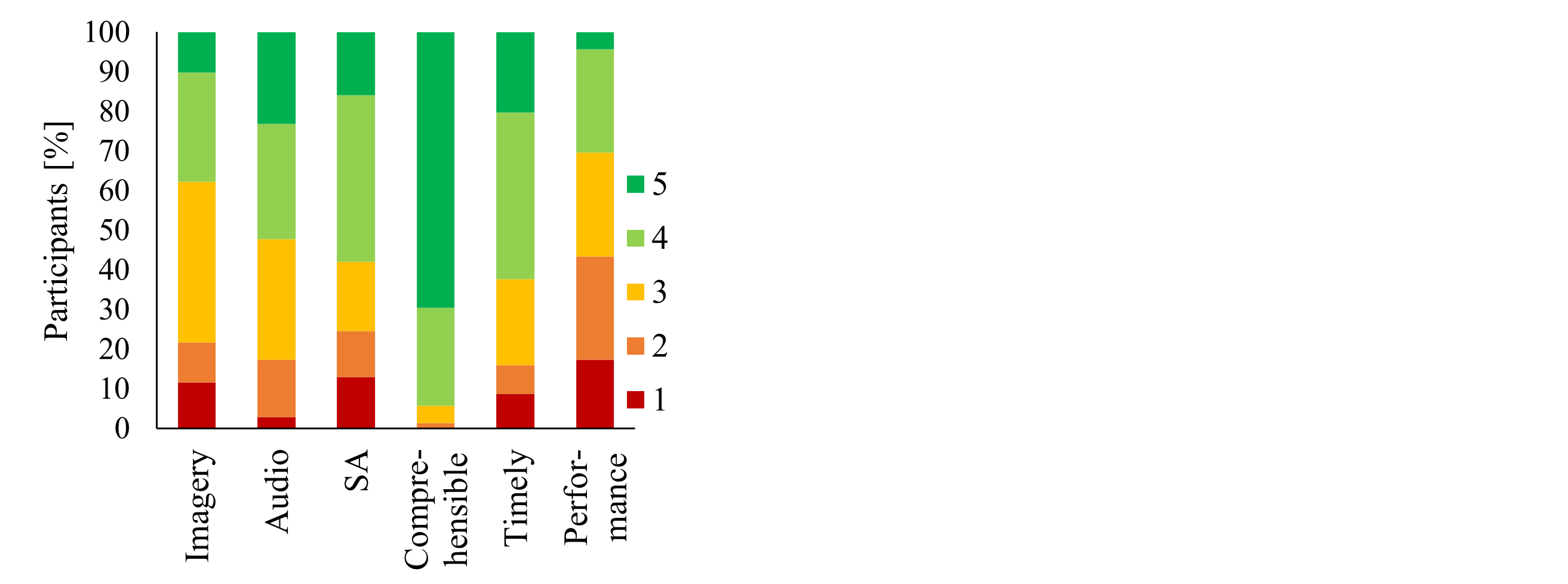}
	\caption{Mid-questionnaire}~\label{fig:mid_que_2020} 
\end{figure}

In the mid questionnaire (Figure~\ref{fig:mid_que_2020}), at least 75 \% of the participants rated the system as helpful, increases SA and warnings are timely in the mid-questionnaire. With regards to performance, ratings were average which corresponds with the ignored warnings indicated by the previously mentioned threshold violations of the experimental group. 

Furthermore, approximately 30 \% of the participants expressed that their reason for ignoring some warnings is due to anticipating the behavior of surrounding vehicles. An example scenario anticipation is the assumption that a b vehicle that is within a safe distance behind the ego vehicle which the driver would not be warned about might change lanes and perform a passing maneuver. In such a scenario, it may also not be safe for the driver to perform a lane change. Thus, they suggested that the warnings should integrate anticipation of the lane change maneuvers of surrounding vehicles to be more acceptable.

\begin{figure}
	\centering
	\includegraphics[width=1.0\columnwidth, trim={0 0 13.6cm 0}, clip]{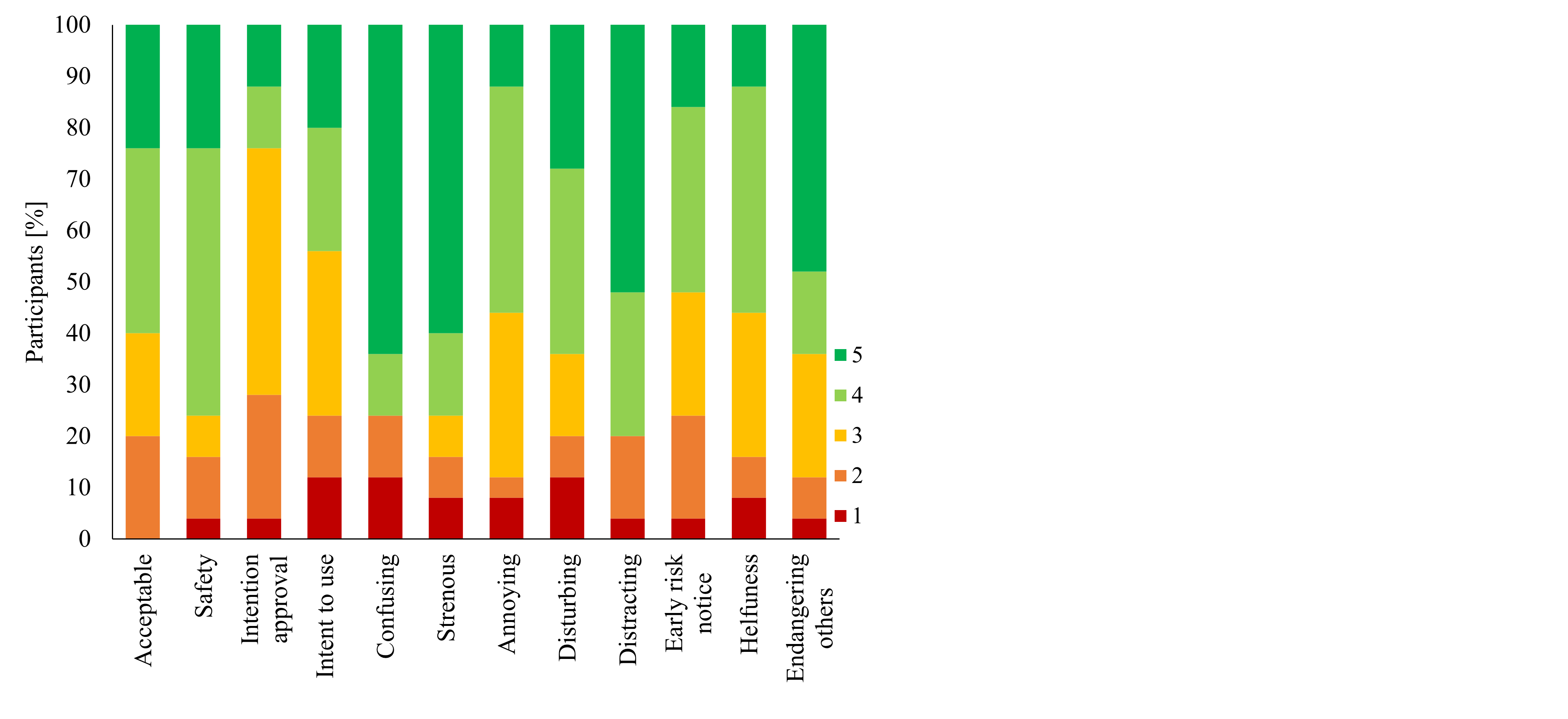}
	\caption{Post-questionnaire}~\label{fig:post_que_2020} 
\end{figure}


In the post questionnaire (Figure~\ref{fig:post_que_2020}), more than 70 \% of the participants rated the system helps to increase SA and is desirable for every day use. The participants also commented that the approval imagery is useful during lane keeping. However, more than 50 \% of the participants (indicated by the corresponding average ratings) commented that the intention approval imagery is not useful for lane changes at the threshold in which they where given. This was because in most cases, they had almost completed the lane change before it appeared.

\subsection{Discussion} 

The results indicate that the effective warning of drivers to elicit the appropriate response is an important multifaceted problem. These include the interface type as well as design, what information is given, time lead as well as the time duration of the information, and driver attentiveness. The interplay between these variables determines the success of the system. If there are too many false alarms, incomprehensive consideration of situational variables or if the information is not comprehensible in the short interval where it is relevant then the driver might habitually mistrust the system and ignore warnings.

The interface serves as the channel through which a warning reaches the driver and its design holds immense control over the information clarity and impact. As indicated in the results, the AR HUD imagery using familiar color schemes, and a minimalist layout enable drivers to identify the direction or location of the potential danger without taking their eyes of the road. More so, the interface design, how the information is given has to be combined with timing factors to elicit driver trust in the system.

Finding the ideal timing of a warning is as crucial as its content. A warning delivered too early may become irrelevant by the time the driver encounters the hazard, while one delivered too late may be insufficient for safe response. The time lead and duration of warnings or other information given have overlapping requirements given the dynamic and demanding task of driving. As seen in the results, the 2 s visual imagery display is suitable for drivers. However, adding two seconds to the warning thresholds to obtain the safe lane change approval thresholds are unsuitable and lead to driver mistrust of the assistance system. Timing algorithms that consider ego vehicle speed and dynamics can optimize the delivery of warnings, ensuring their effectiveness and impact \cite{Yan:2017,Fu:2020}.

Furthermore, addressing the problem of false alarms is essential for building trust and ensuring drivers do not habitually ignore warnings. To make the warnings relevant in this contribution, driver intention is taken into account to warn drivers if their intended lane change actions may result in a collision with another vehicle. Although this resulted in increased safety and warning compliance in some situations drivers still ignored some warnings. Among the reasons given are the need to anticipate the actions of other vehicles especially those of some aggressive vehicles whose cut-in behaviors were too sharp or frequent resulting in increased number of traffic dangers.

One one hand, the additional consideration of anticipating the maneuvers of surrounding vehicles may be addressed with the method proposed by \cite{Williams:2018}. On the other hand, it was necessary to frequently subject the participants to dangerous situations in order to generate warnings and observe the performance of the system. Thus, designing effective driver warnings is a multi-dimensional task that demands a holistic balancing approach integrating well-designed interfaces, concise information, precise timing, and relentless efforts to minimize false alarms.

\section{Summary, conclusion, limitation, and outlook}
\subsection{Summary and conclusion}
This contribution integrates the study of a lane change assistance with warnings based-on predicted driver intention. An individualized driver lane change intention was applied to warn the drivers of eminent danger. The predicted lane change and keep intentions were matched to warnings that were given to drivers in the event of an eminent collision. The results indicate improved performance in the experimental group for lane change to left and right but lane keep. On the other hand, the approval imagery utilized to encourage drivers to make safe maneuvers are useful during lane keeping but not lane change to left and lane change to right. 

\subsection{Limitation and outlook}
The lane change intention assistance system utilized one set of warning thresholds for all the drivers. However, it would be necessary to integrate various warning levels when relevant to further improve the compliance and performance of drivers during lane change to left and keep maneuvers in the future. Furthermore, the study does not include the nature of the hazard such as distinguishing a car from a truck or the recommended course of action for addressing the danger. In the future, a combination of automatic surrounding vehicle lane change anticipation with the intention supervision system can be utilized to improve the warnings.

\section*{Acknowledgment}
The research reported in this paper is partly supported by the Tertiary Education Trust fund (Tetfund) of the Nigerian Government through an Academic Staff Training and Development (AST\&D) Scholarship received by the first author for her Ph.D. study at the Chair of Dynamics and Control, University of Duisburg-Essen, Germany.

\ifCLASSOPTIONcaptionsoff
  \newpage
\fi



\bibliographystyle{IEEEtran}
\bibliography{bibtex/bib/IEEEabrv,bibtex/bib/mybibfile}
%

%

\begin{IEEEbiography}[{\includegraphics[width=1in,height=1.25in,clip,keepaspectratio]{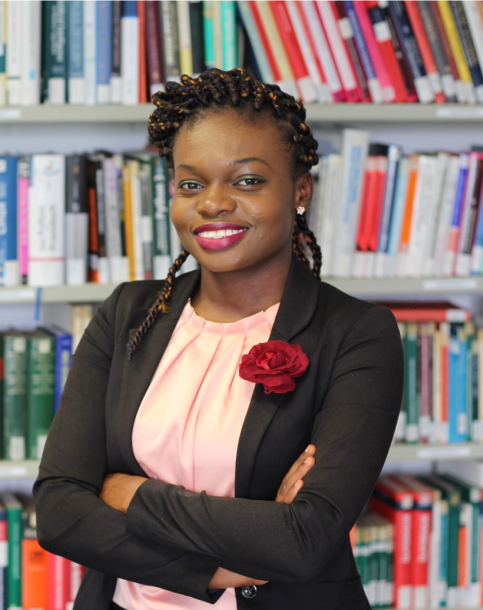}}]{Foghor~Tanshi}~(Member, IEEE) received a Dr.-Ing. degree in Mechanical Engineering from the University of Duisburg-Essen, Germany in 2021. Her current research interests include driver assistance systems and intelligent vehicles, traffic analysis, intelligent transportation systems, human-machine interfaces and interaction, human-machine cooperation, and cognitive systems.
\end{IEEEbiography}


\begin{IEEEbiography}[{\includegraphics[width=1in,height=1.25in,clip,keepaspectratio]{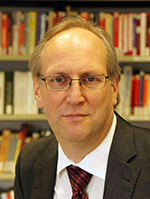}}]	{Dirk~S{\"o}ffker}~(M’10) (Member, IEEE) received a Dr.-Ing. degree
	in Mechanical Engineering and a Habilitation
	degree in Automatic Control/Safety Engineering from
	the University of Wuppertal, Wuppertal, Germany,
	in 1995 and 2001, respectively. Since 2001, he has
	been at the Chair of Dynamics and Control at the
	University of Duisburg-Essen, Germany. His current
	research interests include elastic mechanical structures, modern methods of control theory, human
	interaction with safe technical systems, safety and reliability control engineering of technical systems, and cognitive technical systems.
\end{IEEEbiography}





\end{document}